\documentclass[10pt,twocolumn]{article}
\usepackage[utf8]{inputenc}

\usepackage{graphicx}
\usepackage{tabularx}
\usepackage{authblk}
\usepackage{hyperref}
\usepackage{breakurl}
\usepackage[hang,flushmargin]{footmisc}

\newcommand\blfootnote[1]{%
  \begingroup
  \renewcommand\thefootnote{}\footnotetext{#1}%
  \addtocounter{footnote}{-1}%
  \endgroup
}

\newcounter{protocol}
\newcommand{\protName}{SANS}
\newenvironment{protocol}[1]
  {\par\addvspace{\topsep}
   \noindent
   \tabularx{\linewidth}{@{} X @{}}
    \hline
    \refstepcounter{protocol}\textbf{Protocol \theprotocol} #1 \\
    \hline}
  {\\ \hline
   \endtabularx
   \par\addvspace{\topsep}}

\title{\protName: Self-sovereign Authentication for Network Slices}
\date{}

\author[1,2]{Xavier Salleras}
\author[1,2]{Vanesa Daza}
\affil[1]{Department of Information and Communication Technologies, \par Universitat Pompeu Fabra, Barcelona, Spain}
\affil[2]{CYBERCAT - Center for Cybersecurity Research of Catalonia \par \texttt{\{xavier.salleras, vanesa.daza\}@upf.edu}}

\begin{document}
\maketitle

\section*{Abstract}
\noindent 
5G communications proposed significant improvements over 4G in terms of efficiency and security. Among these novelties, the 5G Network Slicing seems to have a prominent role: deploy multiple virtual network slices, each providing a different service with different needs and features. Like this, a Slice Operator (SO) ruling a specific slice may want to offer a service for users meeting some requirements. It is of paramount importance to provide a robust authentication protocol, able to ensure that users meet the requirements, but providing at the same time a privacy-by-design architecture. This makes even more sense having a growing density of Internet of Things (IoT) devices exchanging private information over the network. In this paper\blfootnote{This work has been accepted to be published in the special issue \textit{Trustworthy Networking for Beyond 5G Networks 2020} of the journal \textit{Security and Communication Networks}.}\makeatother, we improve the 5G network slicing authentication using a Self-Sovereign Identity (SSI) scheme: granting users full control over their data. We introduce an approach to allow a user to prove his right to access a specific service without leaking any information about him. Such an approach is \protName, a protocol that provides non-linkable protection for any issued information, preventing an SO or an eavesdropper from tracking users' activity and relating it with their real identities. Furthermore, our protocol is scalable and can be taken as a framework for improving related technologies in similar scenarios, like authentication in the 5G Radio Access Network (RAN) or other wireless networks and services. Such features can be achieved using cryptographic primitives called Zero-Knowledge Proofs (ZKP). Upon implementing our solution using a state-of-the-art ZKP library and performing several experiments, we provide benchmarks demonstrating that our approach is affordable in speed and memory consumption.

\noindent
\\ \textbf{Keywords:} Self-sovereign Identity, Zero-Knowledge Proofs, 5G Network Slices, Authentication.

\section{Introduction}
\label{sec:introduction}
5G communications enhanced the way how mobile devices are connected to cellular networks. They not solely improved the 4G Radio Access Network (RAN), but also introduced a new paradigm where devices with different specifications are routed through different physical and logical networks, called \textit{network slices}. This opened new business models, for instance, creating network slices for specific services offered by third parties. Like this, a Slice Operator (SO) ruling a network slice may want to offer a service to users meeting some requirements (\emph{e.g.}, users enrolled in a governmental program, users who have paid for using such a service, etc.). Among the growing density of Internet of Things (IoT) devices using 5G communications, we can find examples of devices sharing sensitive data over the network: medical devices exchanging private information or autonomous cars sharing their location with a network slice. Needless to say, this data should not be traced by any SO or eavesdropper. In such a scenario, traditional authentication schemes leak all this data to the SO. As such, Self-Sovereign Identity (SSI) \cite{sovrin} becomes an important feature to implement: systems where users can control, access, and transparently consent their identities, preventing entities from tracking and gathering their personal data. Likewise, the main idea behind SSI systems is to provide a unique mechanism for users to authenticate into different services, providing only the required information, information which shall be non-traceable. 

\textbf{Contributions.} We introduce \protName, a novel self-sovereign authentication approach where a user demonstrates his right to access a service, without leaking any information about him. Our approach is an underlying protocol to be integrated into existing SSI systems, avoiding any user activity to be linked with any other activity done in the past or the future. Moreover, it also prevents the SO or an attacker impersonating him from tracking users' activity. Our protocol grants the user with these main features:

\begin{itemize}
    \item \textbf{Anonymity:} the SO has no way to relate any digital identity with a real identity.
    \item \textbf{Proof of Requirements:} the user can prove that he meets the requirements needed for using a specific service.
    \item \textbf{Non-linkable activity:} the SO has no way to relate any user activity with another activity done in the network.
\end{itemize}

To achieve the aforesaid key features, we use Zero-Knowledge Proofs (ZKPs), cryptographic primitives gaining a lot of momentum in the last years. Since the seminal paper in \cite{gmr85}, demonstrating how ZKPs can prove knowledge of a secret without leaking any information about it, several applications have been envisioned. However, during decades, they were far from being used in real-life applications due to non-existent efficient implementations. Nevertheless, recently, many efficient ZKPs have broken into the scene, revolutionizing not only the state-of-the-art in the area but also the market in scenarios like cryptocurrencies (\emph{e.g.}, Zcash \cite{zcash}) or smart-contracts (\emph{e.g.}, Ethereum \cite{cryptoeprint:2019:191}). Using ZKPs, we can ensure self-sovereign authentication in 5G network slices, as a user would be able to prove his right to access a specific service, requested by an SO, without leaking any information about him.

\textbf{Roadmap.} In Section \ref{sec:background}, we expose the background required to understand the context of the problem and also our solution. In Section \ref{sec:relatedwork}, we introduce the relevant work done concerning our topic. In Section \ref{sec:oursolution}, we present our protocol with all details, including the security analysis, the implementation and the performed experiments. Conclusions are provided in Section \ref{sec:conclusions}.

\section{Background}
\label{sec:background}
In this section, we first introduce the basics of 5G network slicing, and later, for the sake of completeness, we provide an overview of what ZKPs are and how they could be applied to our protocol.

\subsection{5G Network Slicing}

5G is the fifth generation of mobile communications \cite{5gstandards}, which achieves faster speeds than LTE networks and more reliable service. The 5G network is split into different network slices, which are independent networks dedicated and optimized for specific services. This new architecture is built employing Software-Defined Networking (SDN) and Network Functions Virtualization (NFV), along with the physical infrastructure. All these changes lead to higher performance: higher speeds, lower delays, and much less network latency. As depicted in Figure \ref{fig:5garch}, different kinds of User Equipment (UE) are part of different slices, depending on their specifications or the services they are willing to use. In a nutshell, the main network slices are:

\begin{figure*}
  \centering
  \includegraphics[width=380pt]{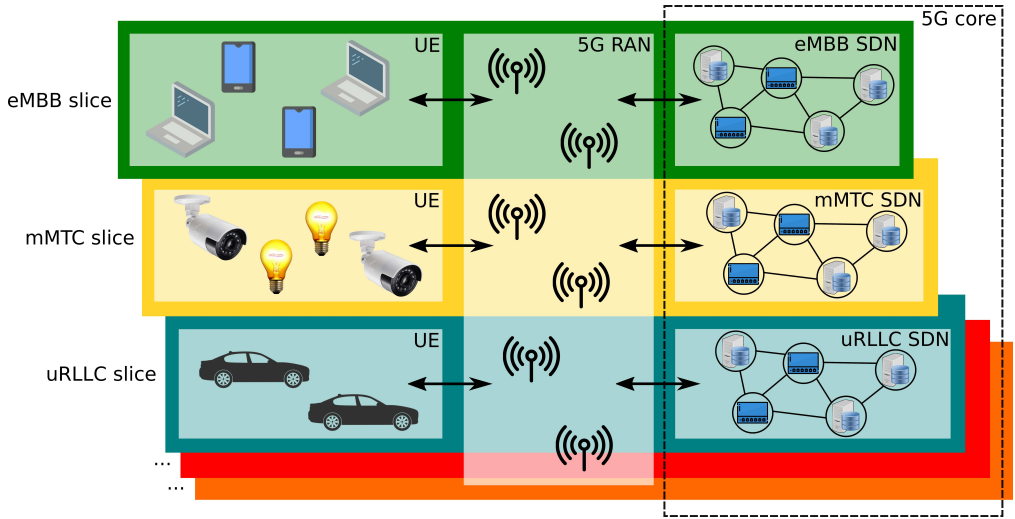}
  \caption{General 5G architecture overview.}
  \label{fig:5garch}
\end{figure*}

\begin{itemize}
    \item \textbf{eMBB slice:} The enhanced Mobile Broadband (eMBB) slice is meant for services that require high bandwidth, like Internet browsing, high definition video streaming, virtual reality, etc.
    \item \textbf{mMTC slice:} The massive Machine Type Communications (mMTC) slice aims to group a high density of devices, which do not have other essential requirements like a low latency or a high bandwidth. Examples of this are IoT devices, specifically in the context of smart cities.
    \item \textbf{uRLLC slice:} The ultra-Reliable and Low-Latency Communications (uRLLC) slice aims to provide very low network latency, a crucial requirement for services like autonomous driving or remote management.
\end{itemize}

As depicted in Figure \ref{fig:5garch}, users connect their UE to the small 5G cells of the 5G RAN, which forward the connections to the 5G core network, split into different software-defined networks (\emph{i.e.}, eMBB, mMTC, uRLLC...).

Furthermore, access to the 5G core network is allowed not solely from the new 5G RAN, but also from other networks like the 4G RAN or optical fiber connections, depending on the requirements of the service. As such, we understand 5G as a heterogeneous network (HetNet), a network interconnecting devices with different specifications and protocols, where a common and trustworthy authentication scheme would be a desirable feature. 

\subsection{Zero-Knowledge Proofs}

A Zero-Knowledge Proof (ZKP) \cite{gmr85} is a cryptographic primitive which allows a prover $P$ to convince a verifier $V$ that a statement is true, without leaking any secret information. In particular, ZKPs must satisfy 3 properties:

\begin{itemize}
 \item \textbf{Completeness:} If the statement is true, $P$ must be able to convince $V$.
 \item \textbf{Soundness:} If the statement is false, $P$ must not be able to convince $V$ that the statement is true, except with negligible probability.
 \item \textbf{Zero-knowledge:} $V$ must not learn any information from the proof beyond the fact that the statement is true.
\end{itemize}

Moreover, $V$ may also be interested in an additional property, the \textit{proof of knowledge}, which guarantees that $P$ knows the secret information about the statement. This secret information that the prover knows is usually called witness $w$. In other words, $P$ wants to prove knowledge of a secret witness $w$ for which a set of operations hold. Such operations are defined by a \textit{circuit}, a graph composed of different wires and gates, which leads to a set of equations relating to the inputs and the outputs of these gates. Each of these equations is called a \textit{constraint}. 

To achieve their goal, both $P$ and $V$ need to interact several times. However, as iterating is not always a desirable property, another kind of ZKPs called Non-Interactive ZKP (NIZKP) \cite{Blum:1988:NZA:62212.62222} arose. In this case, $P$ proves a statement to $V$ by sending him a single message, without interaction. First NIZKP schemes were far from being implemented, due to their impractical computing requirements. Here is where one of the most popular ZKPs arose, zk-SNARKs, which are Zero-Knowledge Succinct and Non-interactive ARguments of Knowledge \cite{cryptoeprint:2013:879}. This kind of proof is short and succinct: it can be verified in a few milliseconds. In this scheme, a trusted setup is required, in order to get some public parameters used either by $P$ and $V$ as a reference to generate and verify proofs. These parameters are called the Common Reference String (CRS). If an attacker was able to get the secret random values $\tau$ used to generate the CRS, he would be able to generate false proofs. For this reason, the initial setup is commonly made through a secure Multi-Party Computation (MPC) \cite{cryptoeprint:2017:1050}, which generates the required parameters using a distributed computation protocol. Therefore, zk-SNARKs are composed of three algorithms: \textit{setup}, \textit{prove} and \textit{verify}. The computing complexity of some of these elements depends on the number of gates $n$, which is the number of operations that we do for proving a specific statement.

There are also other interesting kinds of ZKPs rather than zk-SNARKs, like Bulletproofs \cite{cryptoeprint:2017:1066}. As shown in Table \ref{tab:ev}, Bulletproofs are constructions whose proof size is larger than zk-SNARKs, where the complexity is $O(\log n)$ versus the constant proof size complexity of zk-SNARKs. Moreover, zk-SNARKs are also faster in verifying time complexity. Even when Bulletproofs have linear proving time complexity, the large number of operations required for every constraint leads to a high proving time in practice. As such, the main advantage of Bulletproofs is that they do not require a trusted setup.

Another interesting kind of ZKPs are zk-STARKs (Zero-Knowledge Succinct Transparent ARgument of Knowledge) \cite{cryptoeprint:2018:046}, whose size is much higher than zk-SNARKs and Bulletproofs ($O(\log^2 n)$). One of their main advantages is that like Bulletproofs, they do not require a trusted setup. Another advantage of zk-STARKs is that they are supposed to be post-quantum secure, which is not the case of zk-SNARKs and Bulletproofs.

Regarding the security of the schemes described above, the soundness property of each scheme relies on different security assumptions \cite{assumptions}. As shown in Table \ref{tab:ev}, zk-SNARKs use a strong assumption, the q-Power Knowledge of Exponent ($q$-PKE) assumption, while Bulletproofs or zk-STARKs use better approaches: the Discrete Logarithm Problem (DLP) and Collision Resistant Hash Functions (CRHF), respectively.

\begin{table*}[ht]
\vspace{-0.2cm}
\caption{Comparison of different ZKP constructions, where $n$ is the number of gates of the circuit.}\label{tab:ev}
\vspace{0.1cm}\centering
\begin{tabular}{c|c|c|c|c|c}
\hline
   & \textbf{Trusted setup} & \textbf{Prove}  & \textbf{Verify} & \textbf{Proof size} & \textbf{Assumption} \\
  \textbf{zk-SNARKs} \cite{cryptoeprint:2013:879} & yes & $O(n \log n)$ & $O(1)$ & $O(1)$ & $q$-PKE\\

  \textbf{Bulletproofs} \cite{cryptoeprint:2017:1066} & no & $O(n)$ & $O(n)$ & $O(\log n)$ & DLP\\

  \textbf{zk-STARKs} \cite{cryptoeprint:2018:046} & no & $O(n \log^2 n)$ & $O(\log^2 n)$ & $O(\log^2 n)$ & CRHF\\

  \hline
\end{tabular}
\end{table*}

However, one of the most important improvements regarding ZKPs is the zk-SNARK construction introduced in \cite{cryptoeprint:2016:260}, which introduces the most efficient zk-SNARK designed so far. One of its main improvements is that the verifier has to evaluate a single equation, using only three pairings, instead of five equations and twelve pairings, as done in \cite{cryptoeprint:2013:879}. Such improvements led to a huge usage of this construction in different applications like Zcash.

Another critical research topic is resilience against quantum attacks. An essential contribution regarding this topic has been done in \cite{cryptoeprint:2018:275,cryptoeprint:2018:828}, where a new zk-SNARK constructions believed to be post-quantum secure is introduced.

Regarding the scalability of the implementations, a significant contribution has been done in \cite{cryptoeprint:2019:099}. They propose Sonic, a zk-SNARK construction which requires a trusted setup, but with the difference that such a setup supports different circuits and is also updatable, meaning that the scheme can be continuously improved. As during the setup, a CRS is made public, by using an updatable CRS model \cite{cryptoeprint:2018:280} any user can update the CRS, and he can also prove that it was done correctly, employing a proof of correctness. If this proof is verified, the new CRS can be trusted as long as either the old CRS or the user who did the update was honest. Moreover, zk-SNARK constructions without the need for a trusted setup have also been designed, like the one in \cite{cryptoeprint:2017:1132}. 

Having in mind the schemes described above, the ZKP construction that best fits our solution (at the moment of writing this) are zk-SNARKs. We need proofs to be succinctly verifiable to not overload the verifier and at the same time, it is also preferred to have proofs with a constant size. In that regard, the Groth'16 construction \cite{cryptoeprint:2016:260} provides a reasonably efficient prover, so it could be the preferred option for \protName, as having a construction with efficient proving and verifying algorithms is of paramount importance in our scenario. In Section \ref{subsec:experiments} we show the results of several experiments we have done in this regard. However, we recall the fact that our solution could be used with other ZKP constructions if better options arise.

\section{Related Work}
\label{sec:relatedwork}
Self-Sovereign Identity (SSI) has gained a lot of interest in the last years. The author in \cite{allen} envisioned an SSI system where users can control, consent, and widely use their identities among different services, along with other properties. These properties were redefined in \cite{sovrin} by the Sovrin Foundation\footnote{https://sovrin.org/}. They introduced the guidelines on how SSI systems can be implemented along with blockchain technologies, providing a distributed architecture of trust without central authorities managing users' data. In this regard, SSI authentication schemes like the one proposed in \cite{8489316} make use of blockchain technologies for deploying a decentralized and private authentication system.

A good review of the state-of-the-art regarding this topic is done in \cite{sov1}. As they state, ZKPs allow a user to prove ownership of an identity, \emph{i.e.}, proving knowledge of a secret key related to a public key stored in a blockchain. 

As stated previously, the core of network slicing relies on an SDN-based architecture. In this regard, interesting research is addressed in \cite{8851192}, where a novel authentication scheme preventing multiple types of SDN authentication attacks is introduced. This makes even more sense in the context of a medical cloud sharing sensitive information, a fact that has led to schemes \cite{fang2020physiological} guaranteeing a secure authentication in this scenario.

A more specific use case related to our approach is introduced in \cite{Fedrecheski2020SelfSovereignIF}. They state some of the benefits of SSI for IoT devices, like the fact that the identities of the owners of different devices are stored locally in the devices, rather than on a centralized entity (\emph{i.e.}, the SO in our scenario). As explained by the authors, SSI provides a layered authentication system separating application authentication from channel authentication, where the former handles the trust requirements. This grants a more reliable end-to-end security, where secure communication is established among different protocols.

Among the aforesaid studies regarding SSI, to the best of our knowledge, there are no solutions applied to 5G network slices. In this regard, we propose a solution to integrate SSI into network slices in the next section.

\section{Our solution: \protName}
\label{sec:oursolution}
In this section, we first explain our approach with all the required details. Later we analyze the security of our protocol, its computing constraints, and its benchmarks.

\subsection{Protocol description}

We start with a high-level description of \protName, and later move to a more detailed one: a user willing to join a network slice to use its service may be required to meet some requirements, like having paid a subscription fee. As such, the user is a prover $P$ willing to prove to a verifier $V$, the SO, that he has paid such an amount (the statement). Our protocol accomplishes this purpose. To do so, an important requirement of our protocol is being able to prove knowledge of contracts signed using a given secret key: $P$ must convince $V$ that he knows a contract and its signature, which is verified using a public key. The contract can be a secret value, and still, $V$ must be convinced. In order to be efficient, the used signing algorithms have to be ZKP-friendly, and this means that its operations can be reduced to a low number of constraints. For instance, the Edwards-curve Digital Signature Algorithm (EdDSA) \cite{eddsa} is a fast signing algorithm widely used with zk-SNARKs. Moreover, signature algorithms in zk-SNARKs must be combined with efficient hashing functions as well. One of the most efficient zk-SNARK-friendly hashes to the date is Poseidon \cite{cryptoeprint:2019:458}, which needs 8 times fewer constraints for its circuit than the widely used Pedersen hash.

\begin{figure}[ht!]
  \centering
  \includegraphics[width=230pt]{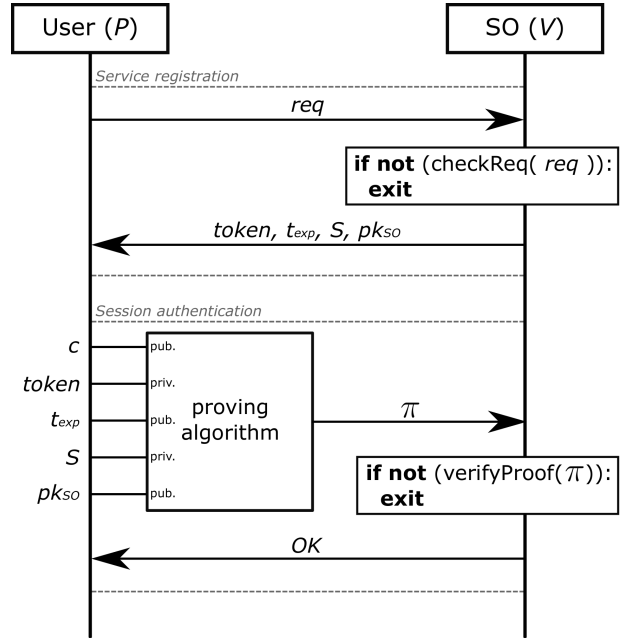}
  \caption{Overview of the service registration and the session authentication.}
  \label{fig:protocol}
\end{figure}

Our authentication scheme is divided into two protocols, depicted altogether in Figure \ref{fig:protocol}. The first one is the \textit{service registration protocol}, to be performed for each issued payment. Its steps are as follows: 

\begin{protocol}{Service registration}
\begin{enumerate}
    \item $P$ provides $V$ some requested information $req$ (\emph{e.g.}, a statement from the bank stating that a payment has been issued).
    
    \item After verifying $req$, $V$ generates a unique byte-array $token$ identifying the user, and sends it to him along with a timestamp $t_{exp}$ representing the contract expiration date. Moreover, $V$ provides a signature $S = sign(token,t_{exp})$ and its public key $pk_{SO}$. 
\end{enumerate}
\end{protocol}

\begin{figure}[ht!]
  \centering
  \vspace{10mm}
  \includegraphics[width=210pt]{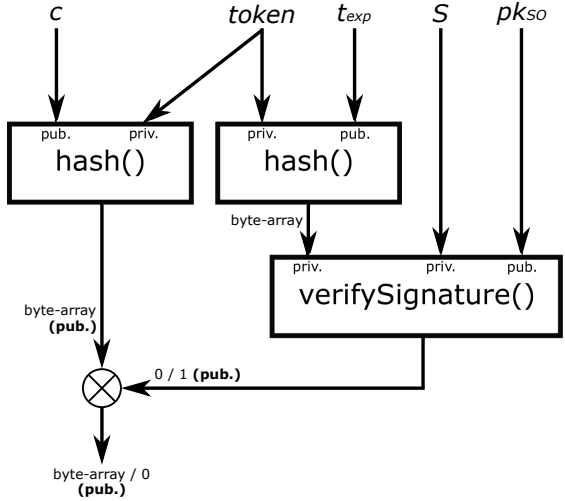}
  \caption{Overview of the circuit used by the the session authentication protocol.}
  \label{fig:circuit}
\end{figure}

After having registered into the service, the user can use the provided parameters to authenticate into the service each time it needs to use it, and thus, create a new session into the service. Moreover, in order to avoid replay attacks \cite{daza2019laser} (\emph{i.e.}, an eavesdropper taking the proof and replying it to the SO), every proof must include the hash of the secret token concatenated to a variable public parameter $c$. Further details of such an approach are discussed in Section \ref{sec:secanalysis}. The \textit{session authentication protocol} is performed as follows:

\begin{protocol}{Session authentication}
    \begin{enumerate}
        \item $P$ computes a proof $\pi$ whose circuit inputs are:
            \begin{itemize}
                \item $c$ (\textbf{public input})
                \item $token$ (\textbf{private input})
                \item $S$ (\textbf{private input})
                \item $pk_{SO}$ (\textbf{public input})
                \item $t_{exp}$ (\textbf{public input})
            \end{itemize}
        \item $V$ verifies the proof $\pi$ and grants the service.
    \end{enumerate}
\end{protocol}

The generated proof $\pi$ proves knowledge of the secret inputs of the circuit depicted in Figure \ref{fig:circuit}. As shown, we prove knowledge of the hash of a secret token concatenated to its expiration date $t_{exp}$. This is our contract, and we also prove that we know its secret signature (signed by $V$) using the public key $pk_{SO}$. This outputs $0/1$ if the signature is verified or not respectively, and this value is multiplied by the output of $hash(c, token)$. If all is correct, the circuit will output a hash, otherwise, the output will be $0$.

\subsection{Security Analysis}
\label{sec:secanalysis}

In this section, we analyze the security of our solution. We also detail how to overcome some possible attacks.

\textbf{False proofs generation.} The main drawback of some ZKP constructions like zk-SNARKs is the need for a trusted setup. In many scenarios, like in Zcash, an untrusty setup could lead to huge losses of money if a malicious party gets the trapdoor $\tau$ and starts to create false transactions. However, this is not a problem in our solution: a different setup can be generated by each SO. If the SO keeps and spreads the trapdoor $\tau$, anyone knowing $\tau$ will be able to access the service by generating false proofs. As such, the protocol is secure as long as the setup is generated only by the SO and he destroys $\tau$. Furthermore, as stated previously, the ZKP construction that best fits our solution at the moment of writing this is the Groth'16 zk-SNARK. As such, the security of \protName{} depends on a $q$-PKE assumption.

\textbf{Elliptic curve attacks.} The security of our solution also relies on the security of elliptic curves. One of the most used curves in ZKPs is a Barreto-Naehrig curve \cite{cryptoeprint:2005:133} called BN128, which security level in practice is estimated to be 110-bits \cite{cryptoeprint:2016:1102}. This means that an attacker willing to break BN128 shall perform $2^{110}$ operations. Other curves like BLS12-381\cite{zcash} estimate around 128-bits of security, with the drawback of heavier group operations. Breaking the security of the used elliptic curve would lead to being able to generate false proofs.

\textbf{Account sharing.} Every computed proof is different since it is generated using random parameters, allowing the user to generate different proofs with the same inputs. As such, the user could generate multiple proofs for other users, which would access the service with a single subscription. To overcome this issue, a simple solution is integrated into our protocol: every proof must include the hash of the secret token concatenated to a variable public parameter $c$. Ideally, this parameter could be a timestamp with a specific accuracy, for instance, the date in format yyyy/mm/dd plus the time in format hh:mm without seconds. Such a hash should be multiplied by the output of the verification of the signature (1 or 0, if verified or not, respectively), and if everything is correct the circuit should output a hash. Like this, an SO receiving the same hash more than once could identify that those proofs have been computed using the same token. As such, if two users are trying to use the service at the very same time, the SO can relate and reject both connections.

\textbf{5G RAN authentication.} One of the main concerns about our solution is to provide a fully private authentication, where the SO cannot learn the identity of the user. In this scenario, we still have another party, the Internet Service Provider (ISP), who acts as an SDN controller providing the architecture and the workflows for optimal network slicing. As such, the ISP learns the identities of the users from the moment that the UE accesses the 5G RAN. To overcome this, we envision the usage of SANS when the UE is required to authenticate for accessing the 5G RAN. In other words, the UE would be proving his right to access the 5G RAN, for instance by proving that the user has paid the last month bill to the ISP.

\subsection{Efficiency Analysis}

This section describes several efficiency considerations of \protName.

\textbf{Computational complexity.} As we saw in Section \ref{sec:background}, the setup protocol depends only on the number of gates, so this protocol has a linear computing complexity $O(n)$. The most consuming operation done by the prover is to compute the coefficients of a polynomial $H(x)$, which can be computed more efficiently employing Fast Fourier Transform (FFT) techniques \cite{7393307}, leading to a computing complexity of $O(n\log{n})$. The verifier has to do a constant computation of group exponentiations and an equation composed of three pairings.

\textbf{Prover optimizations.} There are different operations performed by the zk-SNARK prover which can be parallelized in order to improve its efficiency. This means that CPU and GPU multiprocessing techniques can be applied to speed up the implementations. Even so, the usage of external computing resources as done in \cite{cryptoeprint:2018:691} can be taken into account. For instance, in the case of a prover being a smartwatch with low computing resources, the heaviest computations could be precomputed by the user's phone, whose computing power should be higher.

\textbf{Circuit size.} Our circuit contains a single EdDSA signature combined with two hashes (to the date of writing this, Poseidon seems the best option). The authors of \textit{circomlib}\footnote{https://github.com/iden3/circomlib/} developed optimal EdDSA and Poseidon circuits, which leads our solution to a total size of 7565 constraints and affordable computing times as shown in the next subsection.

\subsection{Implementation and Benchmarks} \label{subsec:experiments}

\begin{figure}[ht]
  \centering
  \includegraphics[width=230pt]{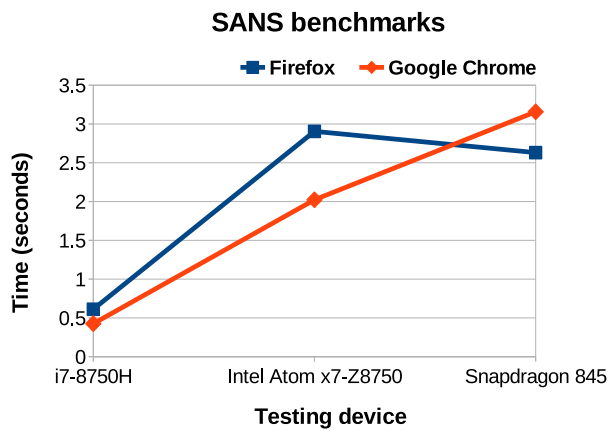}
  \caption{CPUs proving times comparison of SANS using a \textit{snarkjs} implementation executed in multi-core mode for browsers. The used zk-SNARK is Groth'16, and the elliptic curve BN128. Intel Atom x7-Z8750 and i7-8750H run desktop browsers for Linux, Snapdragon 845 runs Firefox and Chrome for Android 10.}
  \label{fig:chart}
\end{figure}

We implemented\footnote{https://github.com/xevisalle/sans} our solution using \textit{snarkjs}, a JavaScript and WASM framework for implementing zk-SNARK applications. The reason for choosing this option is its simplicity for implementing circuits and its portability in web environments. In this regard, we deployed our implementation in a web server, to be executed by different devices using different web browsers. Overall, the number of constraints of this implementation is 7565, and as depicted in the chart of Figure \ref{fig:chart}, our solution outperforms in high-performance CPUs (i7-8750H), either using Mozilla Firefox or Google Chrome. As such, our solution could be used in desktop applications with no problems with regards to performance. 

On the other hand, the proving time increases notably in low-performance processors (Intel Atom x7-Z8750), achieving timings higher than 2 seconds both in Firefox and Chrome. An interesting fact is how Chrome performs slightly better than Firefox in its desktop version, which does not apply to mobile CPUs (Snapdragon 845). Regarding Snapdragon 845, even when it is a top mobile processor, we can see as the results are not as good as i7-8750H. However, the achieved results prove that our solution is feasible in performance, especially when the portability is a priority. Moreover, the memory consumption has been in all tests between 150 and 200MB (not taking into account what is consumed by default by the browsers).

Furthermore, we also tested \textit{libsnark}\footnote{https://github.com/scipr-lab/libsnark}, a well-optimized C++ zk-SNARKs library achieving excellent benchmarks, but with the drawback of not being as portable as other solutions like snarkjs. For instance, as the authors of libsnark state, the library is not well-optimized for ARM architectures (\emph{e.g.} Snapdragon 845), and the BN128 curve is not supported in this architecture.

\begin{figure}[ht]
  \centering
  \includegraphics[width=290pt]{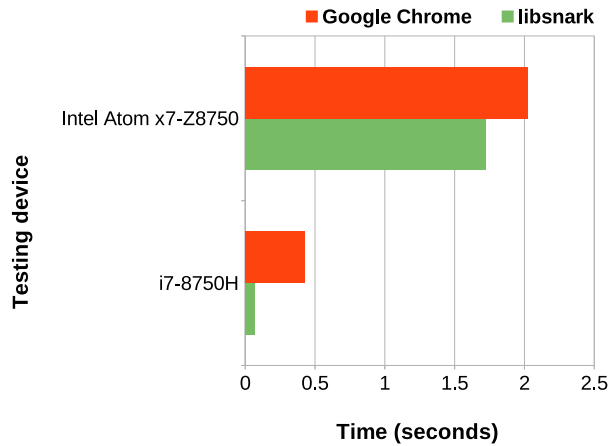}
  \caption{CPUs proving times comparison of \textit{snarkjs} and \textit{libsnark} 7565 constraint circuits executed in multi-core mode. The used zk-SNARK is Groth'16, and the elliptic curve BN128. Intel Atom x7-Z8750 and i7-8750H run a desktop Linux distribution.}
  \label{fig:chart2}
\end{figure}

We implemented a circuit with the same amount of constraints that our solution has, and we executed the prover in multi-core mode using Groth'16 and the BN128 curve. The obtained results are shown in the chart of Figure \ref{fig:chart2}. As can be seen, libsnark achieves much better results than snarkjs, so implementing \protName{} using this library would be even more feasible. Regarding the memory consumption, libsnark performs better as well: around 20 MB in both tested devices. Furthermore, optimized libraries for mobiles and embedded systems would lead to additional performance improvement, so future work in this regard would be an exciting research topic.

\section{Conclusions}
\label{sec:conclusions}
In this paper, we have introduced \protName, a protocol for proving the right of a user to access a specific 5G network slice, without leaking any information about him beyond the fact that he possesses such a right. Our solution is an underlying protocol to be integrated into existing SSI schemes. Moreover, it could be easily extended to other scenarios, like 5G RAN authentication, other kinds of wireless communications, or distributed applications. Even when some ZKP schemes like zk-SNARKs require costly computing operations, we have proved our solution to be affordable in terms of efficiency and memory consumption by implementing \protName{} using existing libraries. Furthermore, we proved the portability of our implementation by testing it on several devices. Nevertheless, future work on optimized ZKP libraries for embedded systems would be interesting, to spread the usage of this protocol.

\section*{Acknowledgements}
The authors are supported by Project RTI2018-102112-B-100 (AEI/FEDER, UE). Moreover, we also want to thank the anonymous reviewers of the special issue \textit{Trustworthy Networking for Beyond 5G Networks 2020} of the journal \textit{Security and Communication Networks}, for their useful comments and suggestions on this work.

\bibliographystyle{unsrt}
\bibliography{bibtex}

\end{document}